\documentclass[]{emulateapj}
\submitted{Accepted for Publication in ApJ}

\newcommand{\Ha }{H$\alpha$}
\newcommand{\HI}{H{\sc i}}
\newcommand{\Dpak } {DensePak}
\newcommand{\kms} {km s$^{-1}$}

\shorttitle{LSB Galaxy Constraints on the NFW Potential}
\shortauthors{Kuzio de Naray, McGaugh, \& Mihos}

\begin{document}
\title{Constraining the NFW Potential with Observations and Modeling of LSB Galaxy Velocity Fields}
\author{Rachel Kuzio de Naray\altaffilmark{1}}
\affil{Center for Cosmology, Department of Physics and Astronomy, University of California, 
  Irvine, CA 92697-4575}
\altaffiltext{1}{NSF Astronomy and Astrophysics Postdoctoral Fellow}
\email{kuzio@uci.edu}
\author{Stacy S. McGaugh}
\affil{Department of Astronomy, University of Maryland, College Park,
  MD 20742-2421}
\email{ssm@astro.umd.edu}
\and
\author{J.~Christopher Mihos}
\affil{Department of Astronomy, Case Western Reserve University,
 10900 Euclid Avenue, Cleveland, OH 44106}
\email{mihos@case.edu}

\begin{abstract}
We model the NFW potential to determine if, and under what conditions, the 
NFW halo appears consistent with the observed velocity fields of low surface 
brightness (LSB) galaxies.  We present mock \Dpak\ IFU velocity fields 
and rotation curves of axisymmetric and non-axisymmetric potentials that are 
well-matched to the spatial resolution and velocity range of our sample 
galaxies.  We find that the \Dpak\ IFU can accurately reconstruct the velocity 
field produced by an axisymmetric NFW potential and that a tilted-ring fitting 
program can successfully recover the corresponding NFW rotation curve.  We 
also find that non-axisymmetric potentials with fixed axis ratios change only 
the normalization of the mock velocity fields and rotation curves and not their
shape. The shape 
of the modeled NFW rotation curves does not reproduce the data: these 
potentials are unable to simultaneously bring the mock data at both small and 
large radii into agreement with observations.  Indeed, to match the slow rise 
of LSB galaxy rotation curves, a specific viewing angle of the 
non-axisymmetric potential is required. For each of the simulated LSB 
galaxies, the observer's line-of-sight must be along the minor axis of the 
potential, an arrangement which is inconsistent with a random distribution of 
halo orientations on the sky.  
\end{abstract}


\section{Introduction}
Cosmologically motivated numerical simulations of cold dark matter (CDM) 
describe very specifically the properties of the dark matter halos that 
should be observed in the universe.  The simulations show that CDM halos 
are cuspy, meaning that the density of the halo, regardless of its mass, 
rises very steeply toward the center  
\citep[e.g.][]{Dubinski94,NFW96,NFW97,Moore,Reed,Diemand}.  The simulations 
also dictate the range of permissible values of halo parameters based on the 
assumed cosmology of the simulations.  The concentration $c$ of a halo, for 
example, depends on the density of the universe at the time the halo forms, 
which in turn depends on the adopted values of $h$, $\Omega_{m}$, 
$\sigma_{8}$, etc. \citep{NFW96,NFW97}.  Due to this intimate connection to 
cosmology, the values of halo parameters are not arbitrary.

The most well-known description of CDM halo behavior is the cuspy NFW halo 
where $\rho \sim r^{-1}$ \citep{NFW96,NFW97}.  The rotation curves of these 
halos are parameterized by two numbers: the concentration, $c$, and a 
characteristic velocity, $V_{200}$.  These two parameters cannot freely vary, 
nor can they vary independently of the other; there is a correlation between 
$c$ and $V_{200}$ \citep[e.g.][]{NFW97,Jing,Bullock01,Wechsler}.  This 
$c-V_{200}$ relation, combined with the cosmological constraints on $c$, 
means that the expected  rotation curve for a CDM halo of a given mass is 
well-determined.

Though the need for dark matter in disk galaxies has long been indicated by 
flat rotation curves \citep[e.g.][]{Rubin,Bosma}, it has been less obvious 
that the dark matter halos are consistent with cuspy CDM halos.  Because low 
surface brightness (LSB) galaxies are thought to be dark matter-dominated 
down to small radii (de Blok \& McGaugh 1996; de Blok \& McGaugh 1997; 
Borriello \& Salucci 2001, but see Fuchs 2003), their kinematics have been 
used as probes of the density distribution of galaxy mass dark matter halos.  
Rotation curves derived from \HI\ velocity fields and long-slit \Ha\ 
observations are frequently consistent with halos having a cored 
$\rho \sim r^{0}$  density distribution 
\citep[e.g.][]{Flores,DMV,dBB,Marchesini,Cote} rather than the steeper profile 
of the NFW halo.  This result has also been supported by rotation curves 
derived from high-resolution two-dimensional velocity fields obtained with 
integral field spectrographs 
\citep[e.g.][]{Chemin04,Gentile05,Simon05,K06,K08}.

In \citet[][hereafter K06 and K08, respectively]{K06,K08}, we presented 
\Dpak\ Integral Field Unit (IFU) \Ha\ velocity fields, rotation curves, and 
halo fits for a sample of 17 LSB galaxies.  We fit both a cored 
pseudoisothermal halo ($\rho \sim r^{0}$) and a cuspy NFW halo 
($\rho \sim r^{-1}$) to the data and found the halo central densities and 
 rotation curve shapes  to be better described by the cored halo model.    
The NFW fits to the \Dpak\ rotation curves were often found to have 
concentrations lower than what is expected for galaxies in a $\Lambda$CDM 
cosmology (see also Gentile et al.~2007, but see Swaters et al.~2003b for a 
different conclusion) and to favor a power spectrum having a lower amplitude 
on small scales (Zentner \& Bullock 2002, McGaugh et al.~2003, K08).  We 
found that the NFW rotation curves specified by the $c-V_{200}$ relation 
(the rotation curves that our galaxies \textit{should} have according to 
$\Lambda$CDM) are much more steeply rising than the 
observed \Dpak\ rotation curves.  In addition, these cosmologically 
consistent halos show a cusp mass excess at the centers of the galaxies, 
indicating that at least two times more mass is expected in the cuspy CDM 
halos than is allowed by the data.     

\defcitealias{K06}{K06}
\defcitealias{K08}{K08}

CDM halos must be both cuspy and follow the $c-V_{200}$ relation defined by 
$\Lambda$CDM.  The density profiles of the \citetalias{K06} and 
\citetalias{K08} data are not well-described by cuspy halos, nor do the 
galaxies fall on the $c-V_{200}$ relation.  These \Dpak\ results are 
consistent with many previous long-slit and \HI\ studies of LSB galaxies 
\citep[e.g.][]{dBMR,Bolatto02,dBB,Swaters03a}, as well as similar \Dpak\ 
studies by \citet{Simon05}.  That different observational techniques (with
different data reduction and analysis procedures as well as sources of error)
lead to similar conclusions suggests that perhaps the discrepancy
between the NFW halo and the observations does not arise at the telescope or
during data analysis, but rather is due to an incorrect assumption
about the specific form of the NFW halo potential.  

Our goal in this paper is to model the NFW halo to determine if, and under 
what conditions, it appears consistent with the observed data.  Starting 
with an axisymmetric NFW potential,  how must it be modified (e.g., 
introduction of an asymmetry) in order to appear consistent with both the 
observed two-dimensional velocity field and the derived rotation curve?   
We construct a model disk galaxy embedded in an NFW halo and then ``observe'' 
it in the same way as we have observed our sample of galaxies with  \Dpak.  
We then compare the mock velocity field and the derived mock rotation curve 
to the real galaxy data.

We adopt a numerical approach to investigating non-axisymmetric halo
potentials because once axisymmetry is broken, the data analysis
becomes much more complicated.  Noncircular motions and asymmetries
are traditionally investigated by doing a higher-order Fourier decomposition 
of the velocity field \citep[e.g.][]{Schoenmakers,Wong}.  We have tried this 
approach, but it was not sufficiently well-constrained for these difficult 
LSB targets to give unique results.  But we do find that useful constraints 
can still be extracted by simulating what is expected to be observed for 
various hypothesized halo potentials.

The paper is organized as follows. In $\S$ 2 we describe the simulations.  
The axisymmetric NFW potential is explored in $\S$ 3. In $\S$ 4 we describe 
the mock velocity fields and rotation curves produced by a non-axisymmetric 
NFW potential with a constant axis ratio.  We determine in $\S$ 5 the 
non-axisymmetric potentials that best describe the observed galaxy data.  
We discuss our results and conclusions in $\S$ 6.  

\begin{deluxetable}{lcccc}
\tabletypesize{\footnotesize}
\tablecaption{Simulated NFW Halo Parameters}
\tablewidth{0pt}
\tablehead{
\colhead{} &\colhead{} &\colhead{} &\colhead{$V_{200}$} &\colhead{$R_{excess}$}\\
\colhead{Galaxy} &\colhead{} &\colhead{$c$}&\colhead{\kms} &\colhead{$\arcsec$}\\
\colhead{(1)} &\colhead{} &\colhead{(2)} &\colhead{(3)} &\colhead{(4)}}
\startdata
NGC 4395 & &8.6 &87 &59\\
DDO 64 & &9.2 &62 &40\\
UGC 4325 & &6.9 &249 &40\\
F583-1 & &8.7 &83 &37\\
F563-1 & &8.4 &101 &13\\
F583-4 & &9.1 &67 &22\\
UGC 5750 & &9.1 &67 &31\\
F563-V2 & &7.9 & 130 &22\\
F568-3 & &8.2 &110 &13\\
\enddata
\tablecomments{Columns 2 and 3 list the NFW halo parameters for the simulated galaxies.  Listed in column 4 are the radii at which the observed \Dpak\ rotation curves and input NFW$_{constr}$ rotation curves overlap in the minimum disk case.  These are the radii out to which the observed and mock rotation curves are compared.}
\end{deluxetable}

\section{Description of Simulations}
$N$-body simulations show that CDM halos of all masses can be described
by the NFW potential \citep{NFW96, NFW97} and its variants
\citep[e.g.,][]{Diemand, Moore, Navarro2004, Reed}.  These cuspy halo
potentials show a steep rise in the mass density toward the center of
the halo.  Most theoretical estimates of the inner slope of the halo
mass density profile are as steep or steeper than that of NFW, so we 
choose to simulate the NFW halo as the conservative case.  If the
NFW potential predicts a dark matter halo with a steeper density profile  
than is allowed by the observed galaxy data, then even more steeply rising
potentials are automatically excluded.  In some formulations 
\citep[e.g.,][]{Navarro2004}, there is no well-defined inner slope,
which continues to roll over to a value that asymptotes to a flatter value
than NFW.  However, this is a small effect at small radii.  The difference
between the original NFW profile and that of \citet{Navarro2004} is
too small to be detected observationally.

In \citetalias{K06} and \citetalias{K08} we defined a constrained NFW halo, 
NFW$_{constr}$.  We required the halo to match the velocities at the outer 
radii of each galaxy by choosing a value of $V_{200}$ which forced the NFW 
velocities to agree with the data points at large radii with the minimum 
requirement of falling within the errorbars of the data.  We then used the 
$c-V_{200}$ relation \citep{NFW97,dBBM} to determine the corresponding 
cosmologically-consistent concentration. This is adjusted to the `vanilla'
cosmology of \citet{Tegmark} by subtracting 0.011 dex in concentration
\citep[see][]{McGaugh03}.   According to $\Lambda$CDM, these 
are the rotation curves that our galaxies should have.  
The chief remaining uncertainty in the normalization of the $c-V_{200}$ 
relation is the power spectrum.  Rotation curves data prefer lower 
$\sigma_8$ \citep{McGaugh07}.

\begin{figure*}[!ht]
\epsscale{0.65}
\plotone{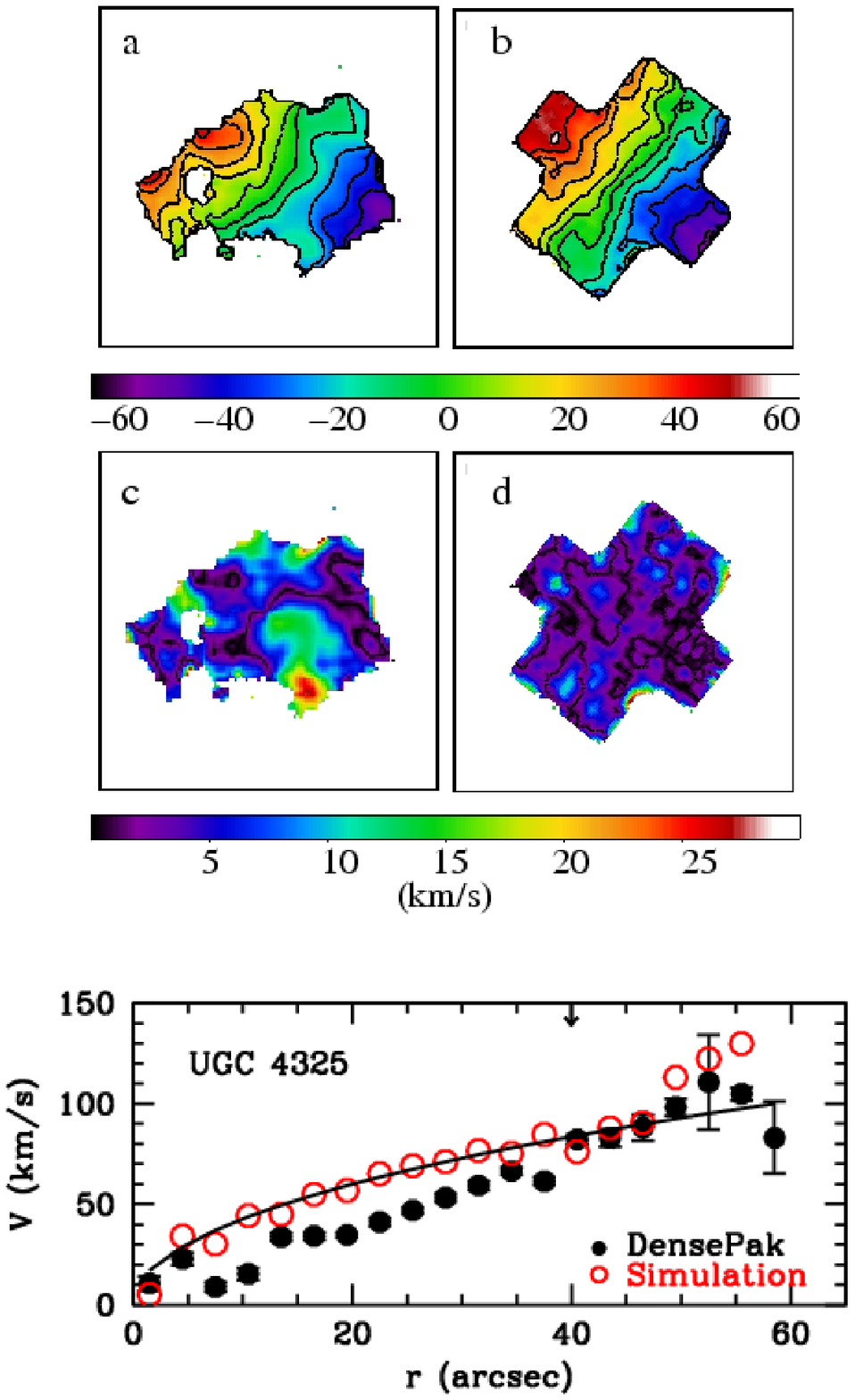}
\caption{\textit{(a)} Observed \Dpak\ velocity field of UGC~4325.  \textit{(b)} Mock \Dpak\ velocity field of the axisymmetric simulation.  Both velocity fields have isovelocity contours at 10 \kms\ intervals.  \textit{(c)} Residual velocity field showing the differences between the UGC~4325 data and the velocity field of the idealized (i.e.~no velocity dispersion), axisymmetric NFW$_{constr}$ halo. \textit{(d)} Same as \textit{(c)} but for the mock \Dpak\ velocity field.  The residuals are large and obvious in \textit{(c)}.  \textit{Bottom:} Observed and mock  rotation curves.  The solid points are the observed \Dpak\ rotation curve of UGC~4325, the solid line is the NFW rotation curve corresponding to the input NFW potential, and the open (red) circles are the rotation curve recovered from the mock velocity field.  The arrow indicates the radius out to which the rotation curves are compared.  The last three points of the recovered mock rotation curve are high because of a lack of fibers at large radii.}
\end{figure*}

\begin{figure*}
\epsscale{0.72}
\plotone{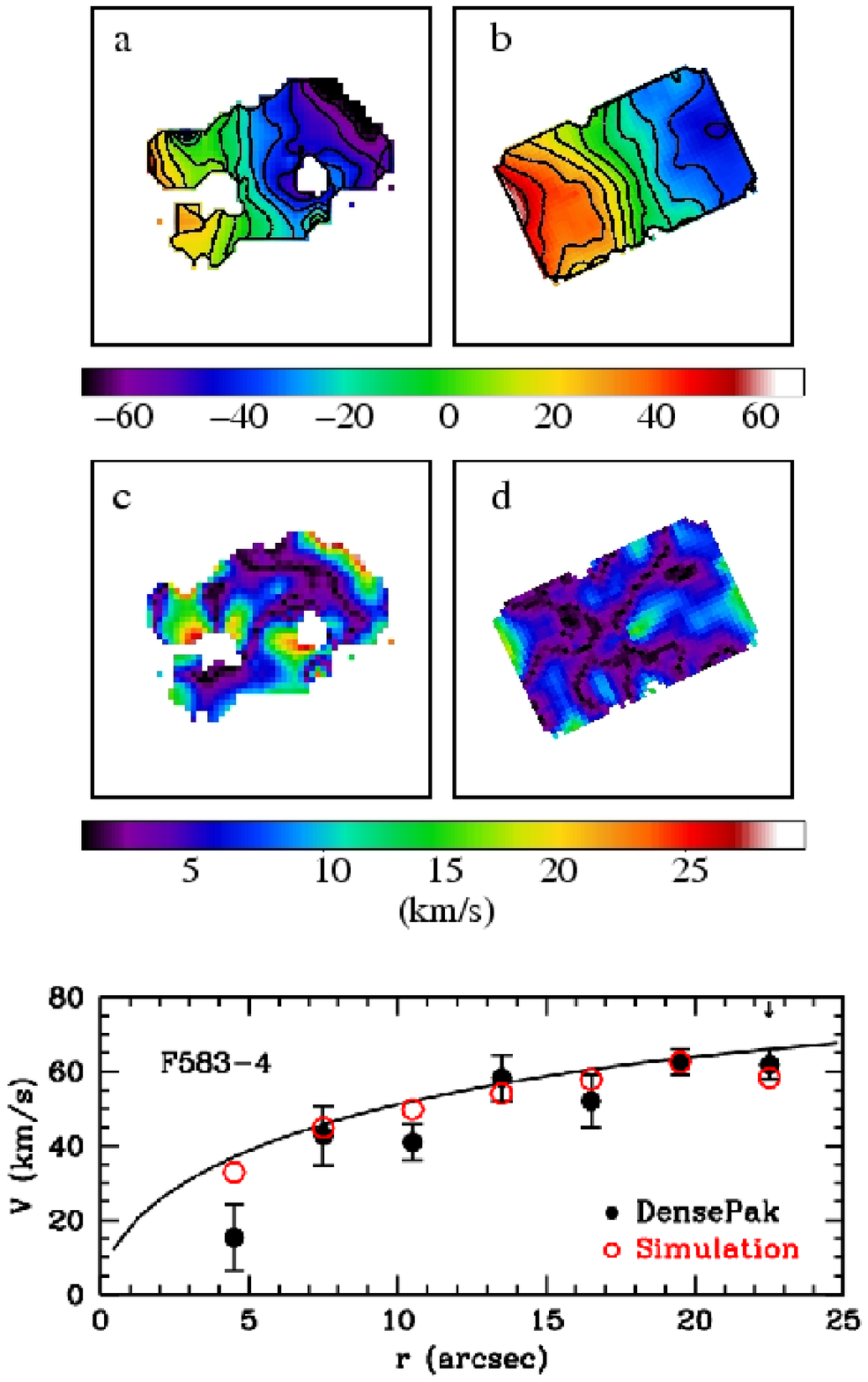}
\caption{Same as Figure 1, but for the less well-resolved galaxy F583-4.}
\end{figure*}

Our goal is to compare mock \Dpak\ velocity fields and rotation curves of 
NFW$_{constr}$ halos to the observed \Dpak\ velocity fields and rotation 
curves of the LSB galaxies in \citetalias{K06} and \citetalias{K08}.  We 
model those galaxies that have well-sampled velocity fields and rotation 
curves that are constrained at large radii by previous long-slit and/or \HI\ 
rotation curves, allowing NFW$_{constr}$ halo fits to be made.  Of our 17 
galaxies, 9 meet these criteria.  The spatial resolution and \Dpak\ coverage 
of these data vary.  The parameters of the NFW$_{constr}$ halos in the limit 
of minimum disk \citepalias{K06,K08} for each of the 9 modeled galaxies are 
listed in Table 1. The galaxies are listed in order of decreasing spatial 
resolution, from NGC~4395 ($\sim$ 20 pc/$\arcsec$) to F568-3 
($\sim$ 375 pc/$\arcsec$).

We developed a code which does a fourth-order Runge-Kutta (RK4) test particle 
integration of point masses moving in a two-dimensional rigid analytic NFW potential.
Specifically, the potential used is:
\begin{equation}
\Phi(R) = -\frac{GM_{200}\ln(1 + \frac{R}{R_{s}})}{Rf(c)},
\end{equation}
where
\begin{equation}
R = \sqrt{x^{2} + (y^{2}/q^{2})}, 
\end{equation}
\begin{equation}
R_{s} = \frac{R_{200}}{c},
\end{equation}
and
\begin{equation}
f(c) = \ln(1 + c) - \frac{c}{1 + c}.
\end{equation} 
In these equations, $M_{200}$ is the enclosed halo mass at radius
$R_{200}$, $q$ is the axis ratio ($q=y/x$), and $c$ is the concentration 
of the halo ($c=R_{200}/R_{s}$).  The halo parameters are set to those of 
the NFW$_{constr}$ halo determined for each galaxy in \citetalias{K06} and 
\citetalias{K08} and are listed in Table 1.  Each simulated galaxy is an 
infinitely thin exponential disk of 10,000 test particles.  For the nearby 
galaxies UGC~4325 and DDO~64, the number of test particles was increased to 
100,000 to ensure sufficient sampling of particles to recreate the higher 
resolution data.  The two-dimensional disk is in the plane of the potential.  
To wash out any numerical pattern noise of the initial conditions, we 
integrate for 50 half-mass rotation periods having 500 timesteps each.  
Each simulated galaxy is given the disk scale length, spatial resolution, 
and inclination of the real galaxy.  

Each simulated galaxy is then ``observed'' by \Dpak.  \Dpak\ is an integral 
field spectrograph on the 3.5 m WIYN telescope at the Kitt Peak National 
Observatory (KPNO).  It is a 43$\arcsec\times28\arcsec$ fixed array of 
3$\arcsec$ fibers with 3.84$\arcsec$ separations.  We model the 85 working 
fibers, as well as the 5 missing or broken fibers, in the main bundle.  
For the galaxies observed in \citetalias{K06} and \citetalias{K08}, the fiber 
bundle orientation on the sky and the total number of pointings per galaxy 
were tailored to each galaxy so that the critical central regions were 
covered by the \Dpak\ fibers.  We aim for obtaining roughly equivalent 
coverage of the simulated galaxies by using similar numbers and alignments 
of \Dpak\ pointings on the simulations.  Prior to extracting a rotation curve, 
these mock \Dpak\ velocity fields are given the velocity dispersion observed 
in the real \Dpak\ galaxy velocity fields.  We have defined the velocity 
dispersion of the \Dpak\ data to be the fiber-to-fiber velocity variation; 
to recreate this in the mock data, we randomly add the desired dispersion to 
the fibers in the mock velocity field.  Rotation curves were then derived 
from the mock \Dpak\ observations by using the NEMO \citep{Teuben} program 
ROTCUR \citep{Begeman}.  ROTCUR treats the observed velocity field as an 
ensemble of tilted rings and fits for the center, systemic velocity, 
inclination, position angle, and rotation velocity in each ring.  The reader 
is referred to \citetalias{K06} and \citetalias{K08} for a more extensive 
explanation of ROTCUR and its application to the \Dpak\ velocity fields.

\section{Axisymmetric NFW Halos}
The most obvious and simple starting point is to assume an axisymmetric halo 
potential.  The axis ratio $q$ is equal to 1 and the test particles move on 
circular orbits.  With this straightforward potential, we can test whether 
or not \Dpak\ observations are sufficient to detect the signature of NFW 
halos in the velocity fields and/or whether the data analysis procedure with 
ROTCUR also suffices to recover NFW rotation curves. 

In Figure 1 we model the NFW$_{constr}$ halo  of UGC~4325 and ``observe'' the 
simulation with 5 pointings of the \Dpak\ array.  The pointings are arranged 
to match the spatial coverage of the real galaxy as much as possible.  
Observed and residual velocity fields for both UGC~4325 and the simulation 
are shown.  The simulated galaxy has the same fiber-to-fiber velocity 
dispersion as the real galaxy: $\sigma$ = 9.0 \kms.  UGC~4325 is one of the 
most nearby ($D\approx 10$ Mpc) and well-resolved galaxies in our sample.  
Diffuse \Ha\ emission was abundant in the galaxy and was detected in almost 
all of the \Dpak\ fibers.

Figure 1 demonstrates two important points.  First, as 
evidenced by the very small residuals of the mock \Dpak\ velocity field, the 
\Dpak\ instrument \textit{is} able to successfully detect an NFW velocity 
field.  The residuals are generally $\lesssim$ 5 \kms\ across the entire 
observed area.  This means that observed \Dpak\ velocity fields are not 
inconsistent with NFW halos because of an inadequacy of the experimental 
design or analysis. Second, the observed \Dpak\ velocity field of UGC~4325 
is not consistent with the axisymmetric NFW$_{constr}$ halo; most of the 
residuals are $\sim$ 10 \kms, and there is a significant region of 
$\sim$ 15 \kms\ residuals near the center.  For UGC~4325, and the other LSB 
galaxies in our sample, 15 \kms\ residuals are non-trivial.  The observed 
fiber-to-fiber velocity dispersions are $\sim$ 6-10 \kms\ \citepalias{K06}, 
and since mass scales as $\sigma^{2}$, the implied mass difference is
a factor of two or more.  In addition, noncircular motions 
caused by disk instabilities, such as spiral or bar modes, are expected to be 
small in LSB galaxies.  The low surface mass densities of the disks  provide
little self-gravity to drive such modes, and 
their high dark matter content provides a higher degree of stabilization
than in high surface brightness galaxies \citep{Mihos97}. 

At the bottom of Figure 1 are the observed and mock rotation curves derived 
from the velocity fields.  The recovered mock rotation curve is consistent 
with the input rotation curve ($\chi^{2}_{r}=0.93$) out to 
$R_{excess}\sim 40\arcsec$, where $R_{excess}$ is defined to be the radius at 
which the observed \Dpak\ rotation curve and the input axisymmetric 
NFW$_{constr}$ rotation curve begin to overlap.  This shows that accurate 
extraction of the rotation curve of an axisymmetric NFW halo with ROTCUR is 
\textit{also} possible.

In Figure 2 we show similar plots for F583-4.  This galaxy has lower spatial 
resolution ($D\approx 49$ Mpc) than UGC~4325 and only a single pointing of 
\Dpak\ coverage.  The velocity fields, residuals, and rotation curves show 
that despite the reduced sampling and lower data quality, an axisymmetric 
NFW halo can still be detected if present.  While the differences between 
the observed and mock velocity field residuals are not as pronounced as in 
the UGC~4325 case, the input NFW rotation curve is successfully recovered by 
ROTCUR ($\chi^{2}_{r}=0.32$).

The velocity field and rotation curve data and simulations plotted in Figures 
1 and 2  together show that the \Dpak\ IFU and the tilted-ring fitting program 
ROTCUR are able to successfully identify an axisymmetric NFW halo in data of 
both high and low quality, if one is present.  That the \citetalias{K06} and 
\citetalias{K08} samples of \Dpak\ observations are inconsistent with NFW 
halos suggests that if the underlying halo potential is NFW, it must not be 
an axisymmetric NFW potential.  This is perhaps not surprising, as CDM 
simulations suggest that the halo potentials are triaxial 
\citep*[e.g.][]{Hayashi07}.   

\begin{figure}
\epsscale{1.0}
\plotone{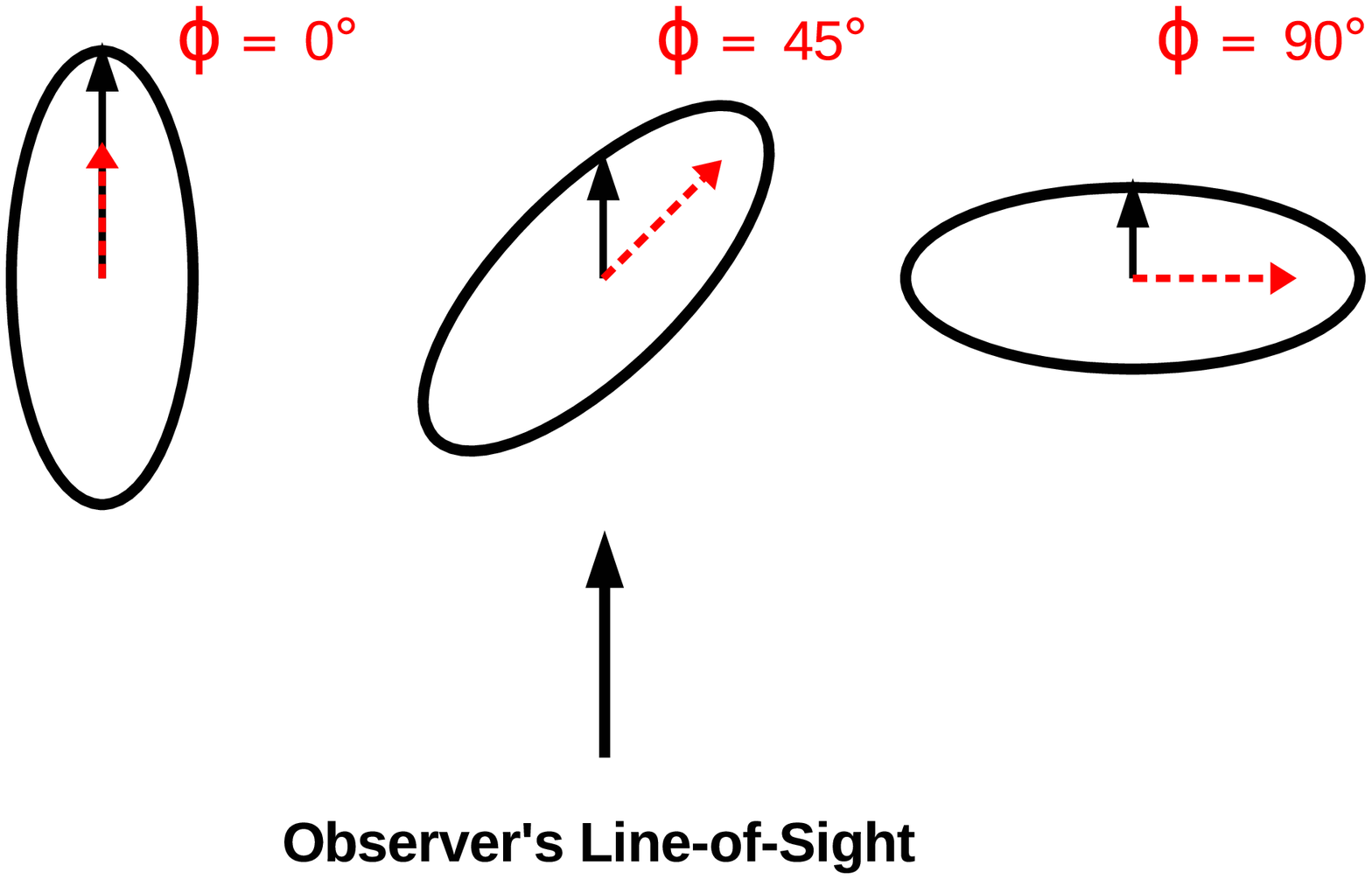}
\caption{Orientation ($\phi$) of the elongated axis of the two-dimensional 
  non-axisymmetric potential with respect to the observer's line-of-sight.  In the $\phi=0\degr$ case, the elongated axis of the potential is along the observer's line-of-sight, whereas in the $\phi=90\degr$ case, it is perpendicular to the observer's line-of-sight. [{\it See the electronic edition of the Journal for a color version of this figure.}] }
\end{figure}

\begin{figure*}
\epsscale{0.55}
\plotone{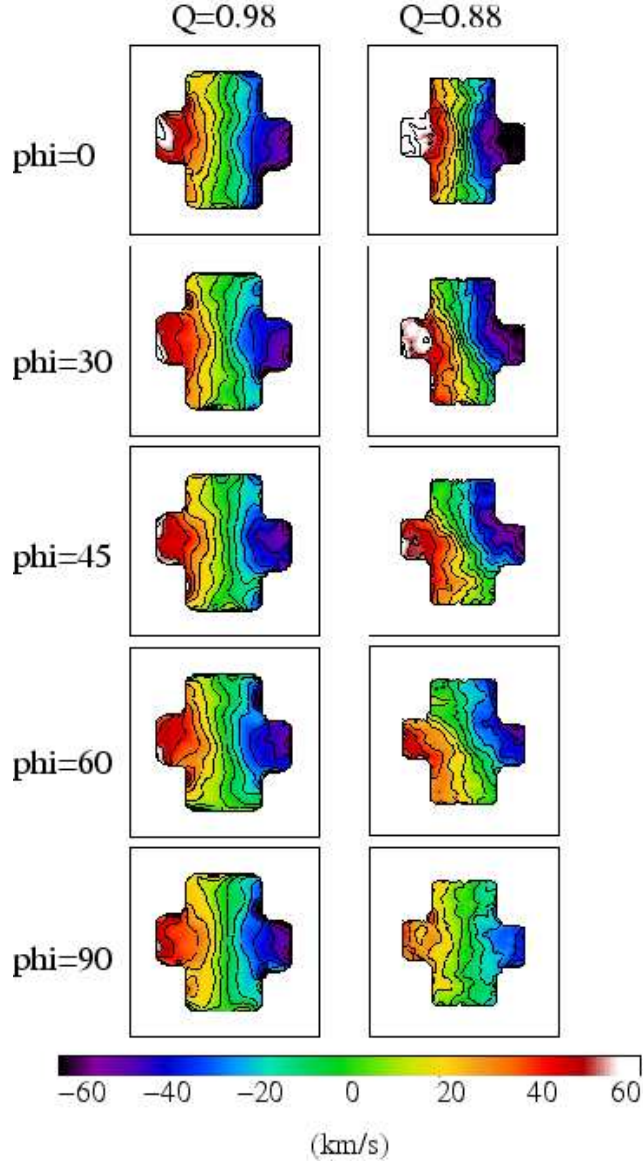}
\caption{Mock \Dpak\ velocity fields for the non-axisymmetric NFW simulations of UGC~4325.  UGC~4325 is a well-resolved galaxy with multiple pointings of \Dpak\ coverage.  Simulations with an axis ratio $q=0.98$ are shown in the left column, and simulations with $q=0.88$ are in the right column. The observer's veiwing angle changes from $\phi=0\degr$ in the top panels to $\phi=90\degr$ in the bottom panels.  Note that the effect of viewing angle is much more pronounced in the $q=0.88$ case.  For easy comparison, all of the velocity fields are on the same color/velocity scale, and isovelocity contours are drawn at 10 \kms\ intervals. }
\end{figure*}

\begin{figure*}
\epsscale{0.52}
\plotone{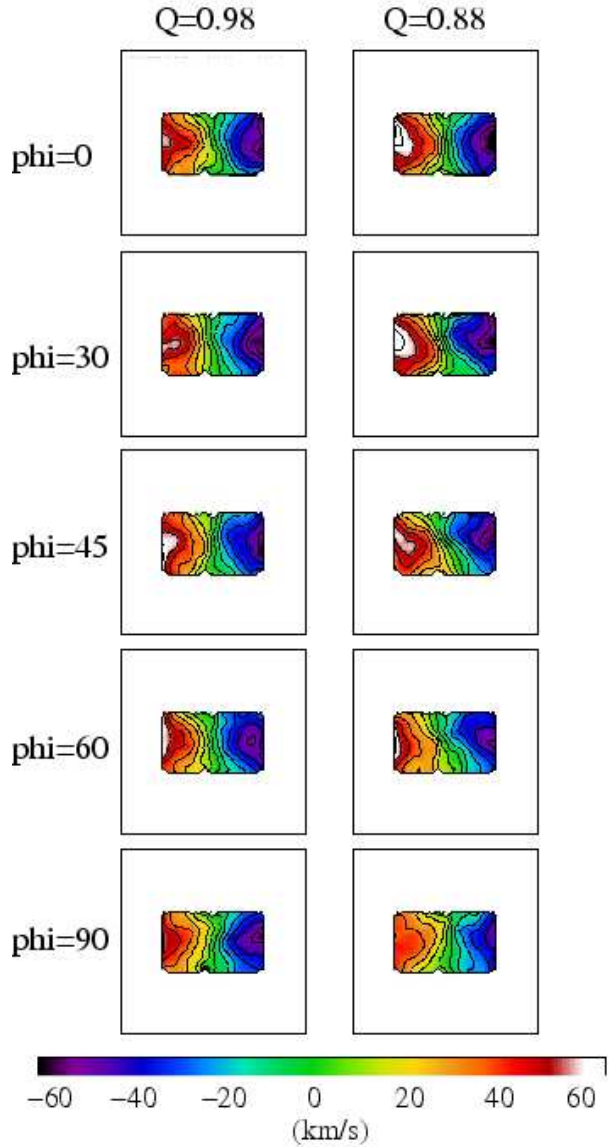}
\caption{Same as Figure 4, but for the mock \Dpak\ velocity fields for the non-axisymmetric NFW simulations of UGC~5750.  UGC~5750 is a more distant galaxy and has only a single pointing of \Dpak\ coverage.  }
\end{figure*}

\begin{figure}
\epsscale{0.87}
\plotone{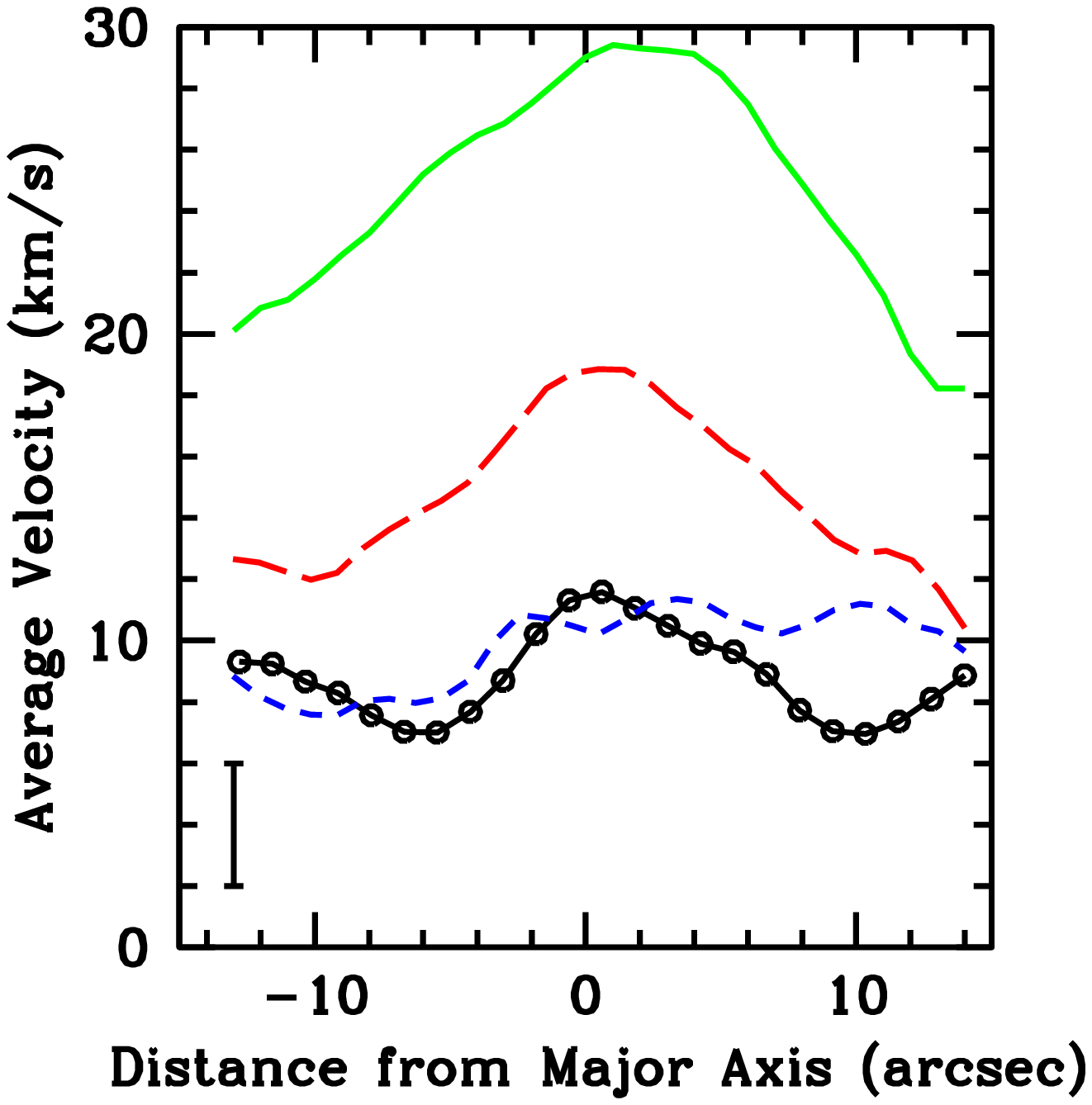}
\caption{A comparison of three mock velocity fields to the observed UGC~4325 data using velocities measured along two slits placed parallel to, and offset from, the minor axis of the velocity fields. The solid (green) line is for the ($q,\phi$) = (0.88,~0$\degr$) mock velocity field, the long-dash (red) line is for the axisymmetric mock velocity field, the short-dash (blue) line is for the ($q,\phi$) = (0.88,~90$\degr$) mock velocity field, and the line+circles are the UGC~4325 data.  At the same distance from the center of the velocity field, the ($q,\phi$) = (0.88,~90$\degr$) mock velocity field has velocities most similar to the observed data.  A typical errorbar is shown in the lower left corner. [{\it See the electronic edition of the Journal for a color version of this figure.}]  }
\end{figure}

\section{Non-axisymmetric NFW Halos with a Fixed Axis Ratio}
We next consider non-axisymmetric two-dimensional NFW potentials with axis 
ratios $q$~$<$~1 that are constant with radius.  These 2D potentials are 
equivalent to 3D prolate dark matter halos in which the long axis of the halo 
coincides with the elongated axis of the disk.  We simulate halos with axis 
ratios $q$ = 0.98, 0.96, 0.94, 0.92, 0.90, 0.88, 0.86, and 0.84, similar to 
the range of non-axisymmetry seen in the CDM simulations of \citet{Hayashi07}. 
 Because axisymmetry has been broken ($q\neq 1$), the test paticles are no 
longer moving on circular orbits and not all lines of sight in the plane of 
the disk are equivalent.  This means that the observed mock \Dpak\ velocity 
field and derived rotation curve are affected not only by the value of $q$, 
but also by the orientation ($\phi$) of the potential's elongation with 
respect to the observer's line-of-sight (see Figure 3).  The potential is 
elongated along the observer's line-of-sight in the $\phi=0\degr$ case, 
whereas in the $\phi=90\degr$ case, the potential is elongated perpendicular 
to the observer's line-of-sight.  For 0$\degr<\phi< 90\degr$, the elongation 
is at an intermediate viewing orientation.   For each value of $q$, the 
orientation of the potential is set to $\phi= 0\degr$, 30$\degr$, 45$\degr$, 
60$\degr$, and 90$\degr$.

\subsection{``Observed'' Mock \Dpak\ Velocity Fields}
In Figures 4 and 5 we show a series of mock \Dpak\ velocity fields for two 
galaxies representative of the range of data presented in \citetalias{K06} 
and \citetalias{K08}:  UGC~4325 and UGC~5750.  As previously mentioned, 
UGC~4325 has high spatial resolution and extended \Dpak\ coverage.  As in the 
axisymmetric case, 5 pointings of the \Dpak\ array are overlayed on these new 
non-axisymmetric simulations.   In contrast, UGC~5750 is a more distant galaxy 
($D\approx 56$ Mpc) and both the real galaxy and the simulations have only 
one pointing of \Dpak\ coverage.  

These are simulations of non-axisymmetric NFW halos that obey the cosmic
$c-V_{200}$ relation.  The virial velocity $V_{200}$ has been chosen to match
each galaxy (the NFW$_{constr}$ halos of \citetalias{K06} and 
\citetalias{K08}).  In this section, we explore the effect of introducing a 
non-axisymmetric potential with equal squashing $q$ at all radii.

In the $q=0.98$ simulations, the potential is nearly circular.  Throughout 
their orbits, the particles maintain a roughly constant distance from the 
center of the potential and as a result, have approximately constant orbital 
speeds.  The viewing angle therefore has little effect on the observed 
velocity field.  The $q=0.98,~\phi=0\degr \rightarrow 90\degr$ mock 
velocity fields appear very similar, looking not only to be consistent with 
different realizations of the same underlying potential, but also very much 
like the mock velocity field of the axisymmetric potential.  

The same cannot be said for the mock velocity fields of the $q=0.88$ 
simulations.  With orbits deviating significantly from circular, a particle's 
orbital speed depends on its location, making the viewing angle quite 
important.  In the $\phi=0\degr$ orientation, the particles moving along the 
observer's line-of-sight are traveling along the long axis of the potential 
and are moving at the maximum orbital speed.  These particles are moving 
faster than particles on circular orbits at the same radius. The 
minimum-maximum velocity range observed by \Dpak\ is larger than what is 
observed in the axisymmetric case, and the derived rotation curves will 
reach higher velocities.  The opposite situation is happening in 
the $\phi=90\degr$ orientation.  In this case, the particles moving along the 
observer's line-of-sight are traveling at the minimum orbital speed and the 
minimum-maximum velocity range observed by \Dpak\ is smaller than what is 
observed in the axisymmetric case.  The rotation curves derived from these 
data will therefore be suppressed.  The difference between the observed 
velocity ranges of the $\phi=0\degr$ and $\phi=90\degr$ velocity fields 
becomes more exaggerated the more noncircular the potential becomes.

Some ($q,\phi$) combinations can automatically be excluded as possible 
descriptions of the observed \Dpak\ galaxy data based simply on the mock 
velocity fields they produce.  The observed \Dpak\ galaxy velocity fields put 
constraints on the allowable velocity range of the mock velocity fields, as 
well as the correlation between velocity and position.  Regardless of how the 
mock rotation curves may turn out, if the observed and mock velocity fields do 
not match, the corresponding simulation is not a viable solution.  For 
example, the ($q,~\phi$) = (0.88,~0$\degr$) mock velocity field of UGC~4325 
shown in Figure 4 can rule out that particular axis ratio/viewing orientation 
combination for that galaxy.  Overall, the mock velocity field covers a much 
larger velocity range than the UGC~4325 data, and when the velocities at the 
same positions in the two velocity fields are compared, they are inconsistent 
over a large portion of the observed area. 

Because of the rapidly rising velocities at the centers of NFW halos, the 
isovelocity contours of NFW velocity fields are pinched in the central regions 
\citep{dBBM}.  This pinch is a distinctive signature of the cuspy NFW halo.  
If the velocity field is noisy or has high velocity dispersion, the 
pinch is more difficult to see.  We can quantify the pinch by measuring 
velocities along slits that are offset from, and parallel to, the minor axis 
of the velocity field.  It is in this fashion that we compare the appearance 
of the UGC~4325 \Dpak\ velocity field and three mock velocity fields.  In 
Figure 6, we have placed a 3$\arcsec \times 28\arcsec$ slit (the width of a 
\Dpak\ fiber and the width of the \Dpak\ array, respectively) 6$\arcsec$ away 
from each side of the minor axis (about the separation of two rows of \Dpak\ 
fibers) of the observed \Dpak\ velocity field of UGC~4325 and the axisymmetric 
NFW mock velocity field (both shown in Figure 1), as well as the 
($q,\phi$) = (0.88,~0$\degr$~and~90$\degr$) mock velocity fields in Figure 
4.  When we plot the average of the measured velocities as a function of 
position along the slit, we can readily see that of the 3 mock velocity 
fields, it is the one produced by the ($q,\phi$) = (0.88,~90$\degr$) 
potential that is most like the data.  At the same position in the velocity 
fields, the axisymmetric, and especially the ($q,\phi$) = (0.88,~0$\degr$), 
mock observations are detecting velocities much higher than the galaxy data. 

In the next section we derive mock rotation curves for all the mock \Dpak\ 
velocity fields and examine the effect that the asymmetry of the potential has 
had on both the normalization and shape of the derived rotation curves.   

\begin{figure*}[!ht]
\epsscale{1.0}
\plotone{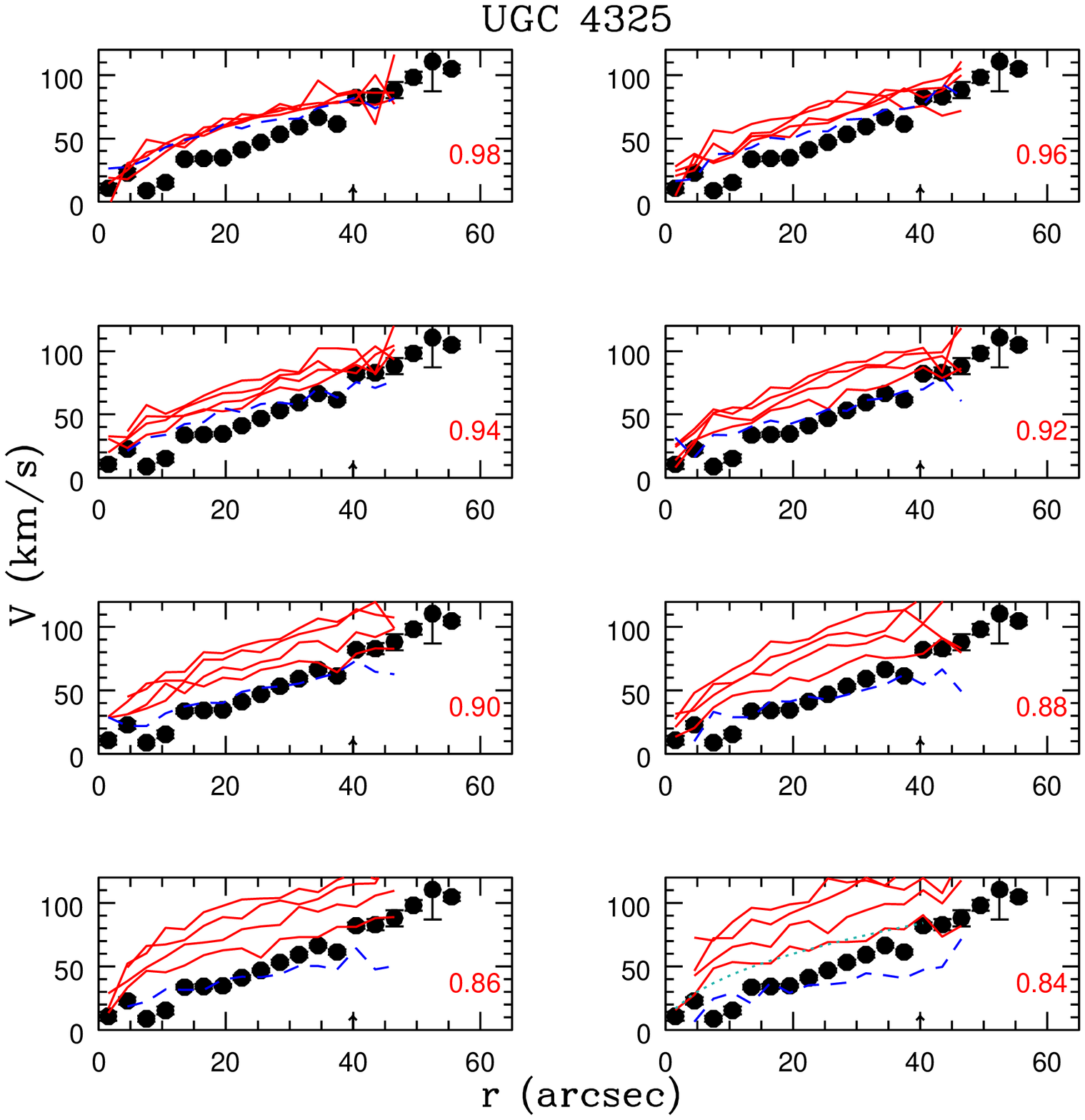}
\caption{Mock rotation curves (lines) for the non-axisymmetric NFW simulations of 
the nearby galaxy UGC~4325 (points).  Each panel is for a different value of the axis ratio $q$ (labeled in the lower right).  The solid lines are for $\phi=0\degr$, 30$\degr$, 45$\degr$, and 60$\degr$.  The dashed line is for $\phi=90\degr$.  For all values of $q$, the $\phi=90\degr$ line has the most overlap with the data at small radii.  The dotted line in the lower right panel is the input NFW$_{constr}$ rotation curve.  The arrows indicate the radius out to which the rotation curves are compared. [{\it See the electronic edition of the Journal for a color version of this figure.}] }
\end{figure*}

\begin{figure*}
\epsscale{1.0}
\plotone{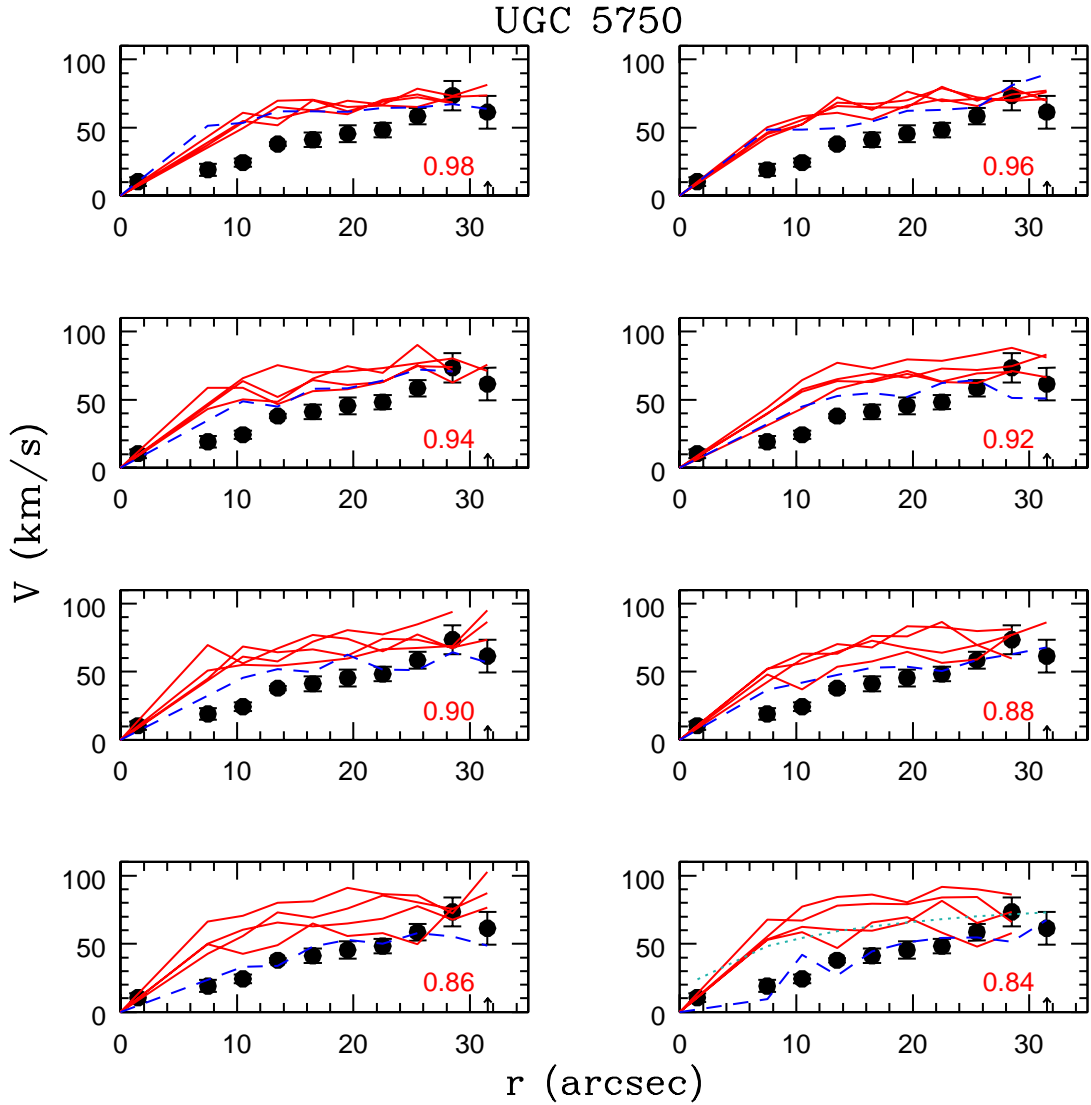}
\caption{Same as Figure 7, but for UGC 5750.  [{\it See the electronic edition of the Journal for a color version of this figure.}] }
\end{figure*}

\begin{figure*}
\epsscale{1.0}
\plotone{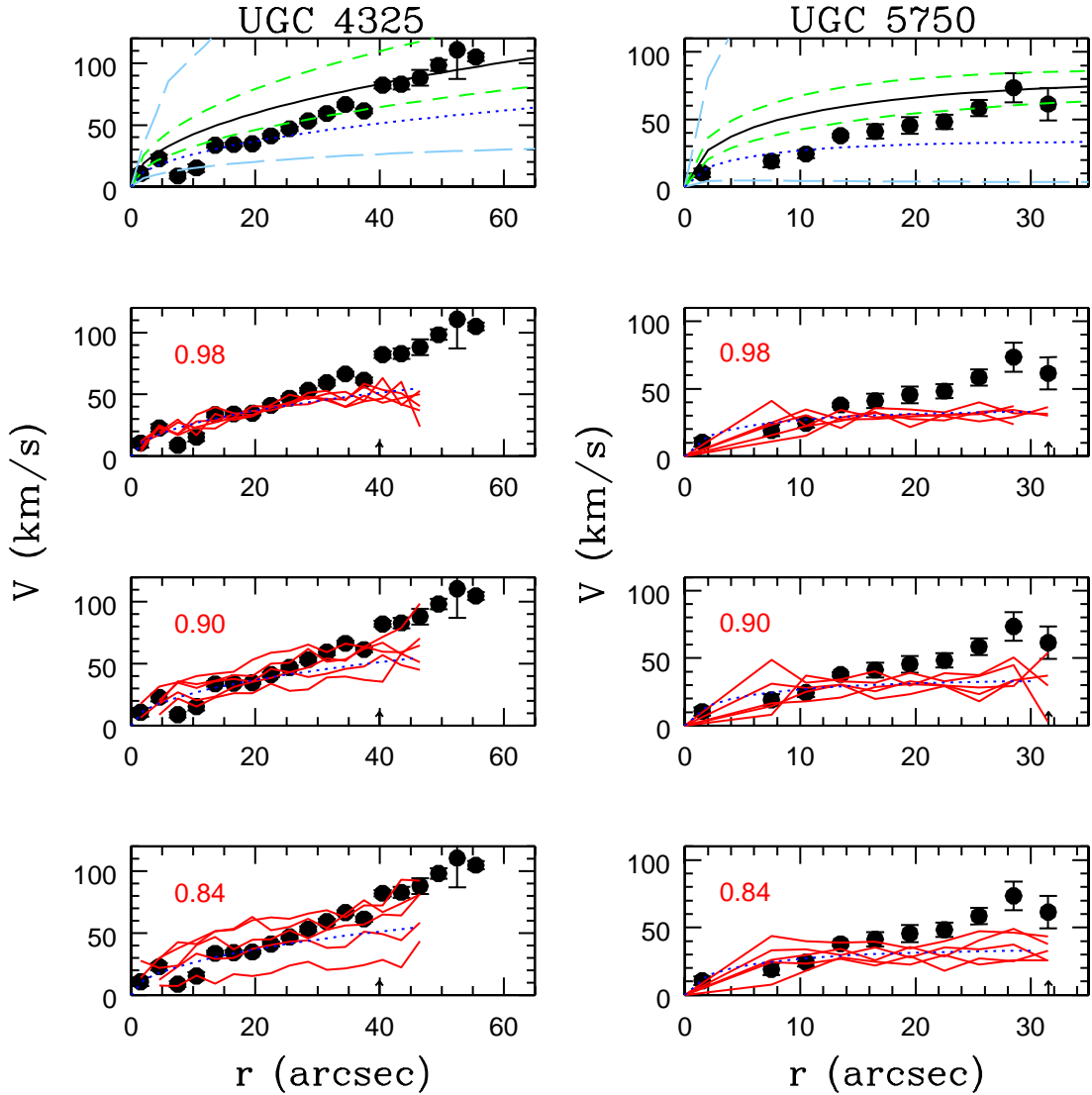}
\caption{\textit{Top Row:} The solid line is the NFW$_{constr}$ rotation curve.  The short dashed lines are the NFW rotation curves corresponding to the $\pm$1$\sigma$ scatter \citep{Bullock01} expected in the concentration in $\Lambda$CDM, and the long dashed lines correspond to the $\pm$1$\sigma$ scatter on $V_{200}$.  The halo of this galaxy could plausibly be drawn from anywhere in this range.  For example, the dotted line is the NFW rotation curve representing a low $V_{200}$ and the lowest corresponding concentration within the scatter.  This halo provides a good initial match to the inner data at the expense of falling well short of the outer data.  \textit{Lower panels:} Mock rotation curves for the non-axisymmetric NFW simulations of this low $V_{200}$ halo (dotted lines).  Each panel is for a different value of the axis ratio $q$ (labeled in the upper left).  The solid lines are for $\phi=0\degr$, 30$\degr$, 45$\degr$, 60$\degr$, and 90$\degr$. The arrows indicate the radius out to which the rotation curves are compared. [{\it See the electronic edition of the Journal for a color version of this figure.}] }
\end{figure*}

\begin{figure}
\epsscale{1.0}
\plotone{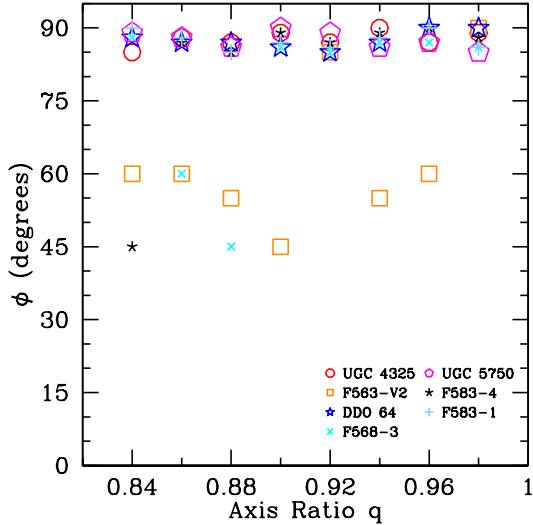}
\caption{The viewing angle $\phi$ that minimizes the differences between the mock and observed galaxy rotation curves out to $R_{excess}$ for each value of $q$ for each galaxy.    Nearly all of the galaxies fall on top of each other in the $\phi=85\degr$ to 90$\degr$ range for all values of $q$.   The optimal viewing angle for F563-V2 (open squares) is  $\phi \sim 55\degr$; there is a stellar bar at this position angle.  The results for F563-1 and NGC~4395 are not shown because $\phi$ is unconstrained due to the radial extent of the data. Nevertheless, the fact that we detect the bar in F563-V2 is an encouraging confirmation of the method. [{\it See the electronic edition of the Journal for a color version of this figure.}] }
\end{figure}

\begin{figure}
\epsscale{1.0}
\plotone{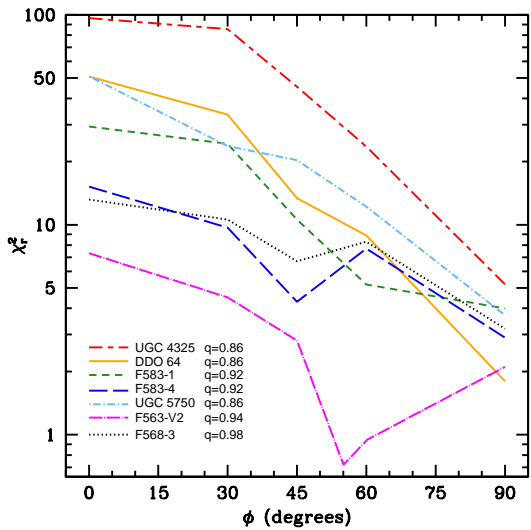}
\caption{The value of $\chi^{2}_{r}$ as a function of $\phi$ for the best-fitting $q$ for each galaxy. Although the $\phi \rightarrow 90\degr$ mock rotation curves are not formally good fits to the observed data, they are the minimum in $\chi^{2}$-space. The exception is F563-V2, for which $\chi^{2}_{r} \approx 1$ for the optimal viewing angle of $\phi\approx55\degr$. [{\it See the electronic edition of the Journal for a color version of this figure.}] }
\end{figure}

\begin{figure*}
\epsscale{0.80}
\plotone{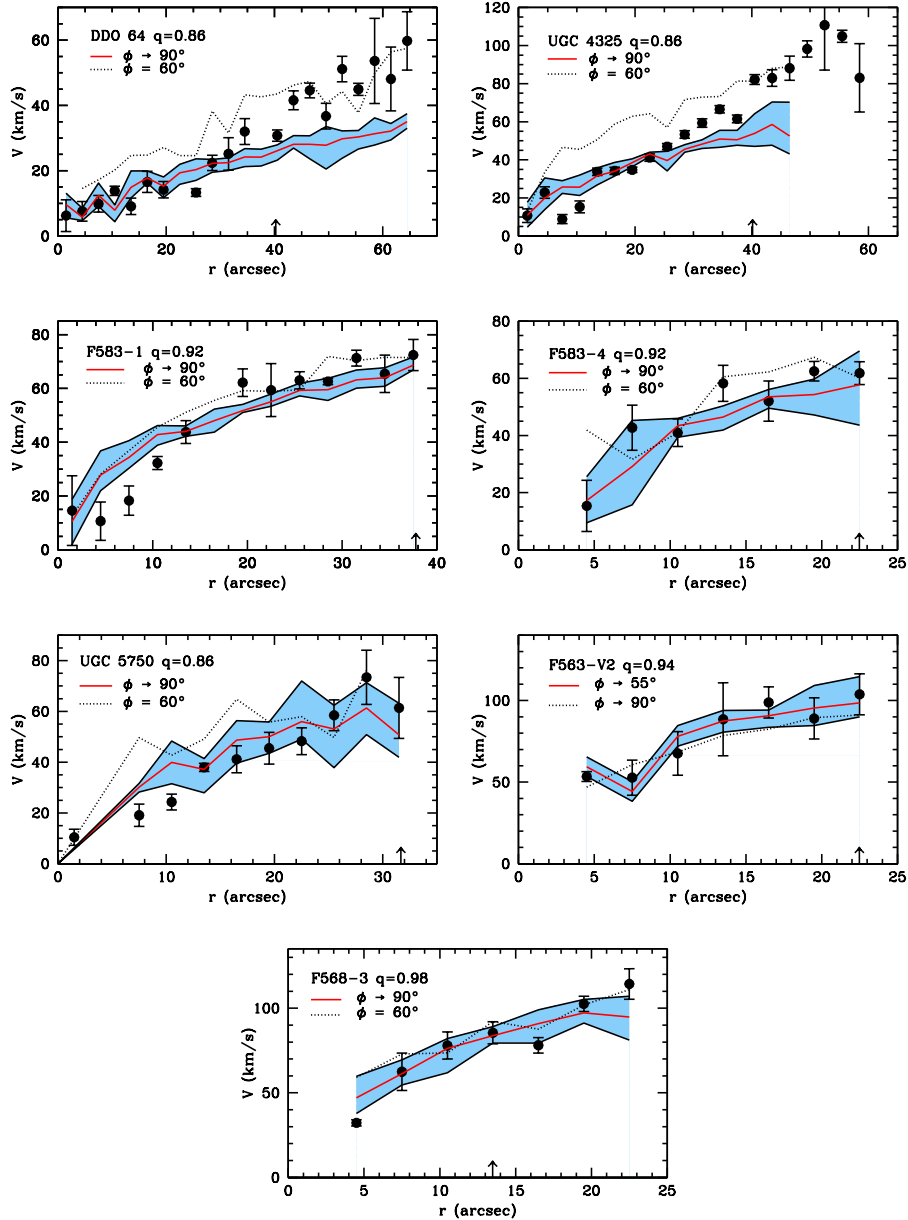}
\caption{The ``best-fitting'' $\phi \rightarrow 90\degr$ rotation curves ($\phi \rightarrow 55\degr$ for F563-V2) for each galaxy.  The solid (red) line is the average of the $\phi=85\degr \rightarrow 90\degr$ ($\phi=45\degr \rightarrow 60\degr$) rotation curves and the shaded (blue) band outlines the spread in those rotation curves.  For comparison, the dotted line is one realization of the $\phi=60\degr$ ($\phi=90\degr$) rotation curve.  The arrow indicates $R_{excess}$.  From top to bottom, the galaxies are ordered by increasing distance.  The simulations of UGC~4325 do not extend to the outermost observed rotation curve point.  The inner 7$\arcsec$ of the mock rotation curves of UGC~5750 are poorly sampled, as in the real data.  The results for F563-1 and NGC~4395 are not shown because $\phi$ is unconstrained. 
[{\it See the electronic edition of the Journal for a color version of this figure.}] }
\end{figure*}

\begin{figure}
\epsscale{0.72}
\plotone{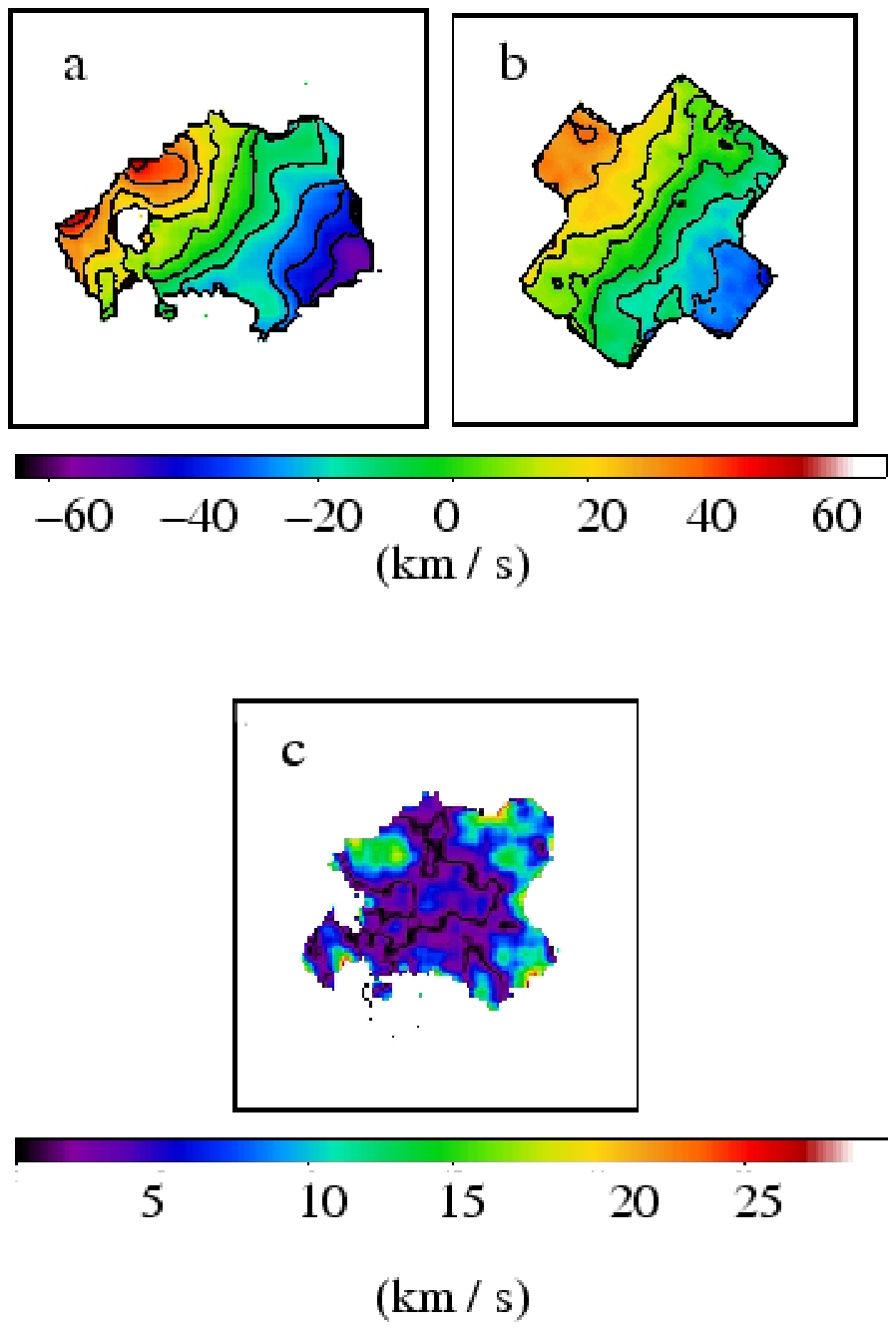}
\caption{\textit{(a)} Observed \Dpak\ velocity field of UGC~4325. \textit{(b)} The ``best-fitting'' non-axisymmetric $\phi=90\degr$ mock NFW velocity field of UGC~4325.   
\textit{(c)} Residual velocity field showing the difference between the observed and mock velocity fields.  The velocity fields are similar in the central regions, but become more mismatched at large radii.}
\end{figure}

\subsection{Derived Mock Rotation Curves}
The LSB galaxies observed with \Dpak\ in \citetalias{K06} and 
\citetalias{K08} surely have some level of noncircular motions, but we have 
assumed only circular motion when deriving the rotation curves with ROTCUR;  
we treat the mock observations the same way.  The test particles are, 
by construction, no longer on circular orbits in these non-axisymmetric 
potentials, but because we did not correct for this in the real data,
we do not correct for it in the mock observations.  Any errors in the
galaxy rotation curves which may have resulted from the assumption of
circular motion will be reproduced in the rotation curves of the mock
observations.

In Figure 7 we show the mock rotation curves for the non-axisymmetric 
simulations of the well-resolved galaxy UGC~4325.  Each panel shows the 
observed rotation curve of UGC~4325 along with the $\phi=0\degr$, 30$\degr$, 
45$\degr$, 60$\degr$, and 90$\degr$ mock rotation curves for a single value 
of $q$.  Similar to the trends seen in the mock velocity fields, we find that 
as the axis ratio decreases and the potential becomes increasingly more 
elongated, the influence of the viewing angle on the inferred rotation curves 
becomes more striking:  the mock rotation curves for the different values of 
$\phi$ spread farther out in velocity space.  We can quantitatively measure 
and compare the shapes of these rotation curves using the ratio of the radii 
containing 80\% and 50\% of the velocity at $R_{excess}$.  The average values 
of R$_{80}$/R$_{50}$ for the UGC~4325 $q=$ 0.98, 0.90, and 0.86 mock rotation 
curves, for example, are 2.7$\pm$0.7, 2.8$\pm$0.8, and 3.1$\pm$0.7, 
respectively.   \textit{This indicates that the overall shapes of the mock 
rotation curves are not changing significantly as $q$ or $\phi$ change; it is 
the normalization of V(r), including V$_{max}$, that is shifting up or down.}  
This is an important point to recognize, as it means that simply adopting a 
different (lower) value of $V_{200}$ for the underlying NFW halo (which 
cannot be done without ignoring cosmological constraints) will not reconcile 
the observed and mock rotation curves; the mock NFW rotation curve will still 
not fit the data properly.   Though none of the mock rotation curves match 
the shape of the entire observed rotation curve of UGC~4325, those that are 
most consistent with the data at small radii are the $\phi=90\degr$ mock 
rotation curves.  This is true for all of the values of $q$ that were 
simulated, but it is the $\phi$ = 90$\degr$ rotation curves in the 
$q\lesssim0.90$ simulations that have the most overlap with the data.      

As a comparison to the well-resolved observations of UGC~4325, we show in 
Figure 8 similar plots of the mock rotation curves for UGC~5750.   Despite 
the lower spatial resolution, we find the mock rotation curves of UGC~5750 to 
behave in very much the same way as the mock rotation curves of UGC~4325.  
\textit{We again see that the amplitude, not the shape, of the rotation 
curve changes as $q$ and $\phi$ change, with the magnitude of the change 
becoming more pronounced as the potential becomes more asymmetric.}  We 
also find that the $\phi=90\degr$ mock rotation curves are again the closest 
to approaching the data, though substantial overlap at small and intermediate 
radii does not occur until $q\lesssim 0.86$.

We examine in Figure 9 how changing the input NFW halo parameters affects the 
mock rotation curves, specifically exploring if $q$ and $\phi$ can change the 
\textit{shape} of a more slowly rising input NFW rotation curve.  For both 
UGC~4325 and UGC~5750, we have simulated NFW halos with low values of 
$V_{200}$ and the lowest corresponding concentration within the scatter of 
\citet{Bullock01} (UGC~4325: $c=5.2$, $V_{200}=140$; UGC~5750: $c=6.9$, 
$V_{200}=60$).  These new input NFW rotation curves fall between the rotation 
curves representing the -1$\sigma$ scatter on $c$ and $V_{200}$ of the 
original NFW$_{constr}$ rotation curves.  In addition, these slowly rising 
rotation curves overlap the observed \Dpak\ data at small radii, in contrast 
to the NFW$_{constr}$ rotation curves which match the data at large radii.
Essentially, we choose to match the inner rather than outer velocities with 
halos drawn from the favorable edge of the plausible cosmological distribution.

For both galaxies, we find the new mock rotation curves to behave similarly to 
the mock rotation curves in Figures 7 and 8.  As the axis ratio $q$ decreases, 
the mock rotation curves for the different values of $\phi$ scatter about the 
input NFW rotation curve, spreading farther out in velocity space.  Even 
though these new mock rotation curves overlap some of the observed \Dpak\ data 
at small radii, none are formally acceptable fits ($\chi^{2}_{r}$ $\gg$ 1).  
More importantly, the shapes of the mock rotation curves are not significantly 
changing as $q$ and $\phi$ change: the average values of R$_{80}$/R$_{50}$ 
for the UGC~4325 $q=$ 0.98, 0.90, and 0.84 mock rotation curves are 
2.3$\pm$0.3, 2.8$\pm$0.4, and 3.1$\pm$1.2.  The mock UGC~5750 rotation curves 
are essentially flat between $\sim$10$\arcsec$ and $\sim$30$\arcsec$, 
preventing useful measurements of R$_{80}$/R$_{50}$.  From Figure 9 we can see 
that regardless of the input NFW halo parameters, $q$ and $\phi$ change only 
the normalization, not the radial behavior, of the mock rotation curves.

It is also worth stressing that, given the behavior of the mock rotation 
curves in Figures 7 and 8, as well as Figure 9, observers should see rotation 
curves with a range of normalizations: there should be rotation curves both 
above and below the nominal rotation curve expected from the $c-V_{200}$ 
relation (compare the mock rotation curves to the dotted lines in Figure 9 and 
the lower right panels of Figures 7 and 8).  But this is, in fact, not what is 
observed in long-slit data \citep{MRdB,dBB}.  LSB and NFW rotation curves 
nearly always differ in the sense that the observed rotation curve velocities 
at small radii must increase so that the data match the models, or 
equivalently, the NFW rotation curve velocities must decrease so the models 
match the data.  LSB rotation curves which are higher than NFW rotation curves 
are seldom found, if ever.  

There is a trade-off between $q$ and $\phi$ such that different combinations 
of the two parameters can produce similar mock rotation curves.  In the 
following section, we explore what combination of ($q,\phi$) minimizes the 
differences between the NFW halo and the \Dpak\ galaxy observations.  We 
simulate the NFW$_{constr}$ halos rather than low $V_{200}$ halos like those 
in Figure 9 because the constrained halos were required to match the 
velocities at the outer radii of each galaxy, a reasonable constraint since 
dark matter must explain the high velocities at large radii where the 
contribution of the baryons has fallen off.  Parameter space is too large 
to explore all possible initial halos.  However, given that plausible 
combinations of NFW $c$ and $V_{200}$ parameters give rather degenerate 
rotation curves, and that $q$ and $\phi$ affect only the normalization, not 
the shape of $V(R)$, our choice should lead to fairly general results.

\section{Minimizing the Cusp Mass Excess with $q$ and $\phi$}
In \citetalias{K08} we showed that there is a substantial cusp
mass excess near the centers of the galaxies when the NFW$_{constr}$
halo is used to describe the dark matter halo.  Evaluating the difference 
between the NFW$_{constr}$ rotation curve and the observed galaxy rotation 
curve in terms of mass rather than velocity, we determined that interior to 
the radius where the two rotation curves begin to overlap ($R_{excess}$), NFW 
halos are at least twice as massive as the galaxy data will allow.  In this 
section, we are interested in determining for each \Dpak\ galaxy what 
combination of ($q,\phi$) minimizes the differences between the observed 
and mock rotation curves out to $R_{excess}$ where, in the limit of zero 
stellar mass, the cusp mass excess is  $\sim 0$ (see Table 1). 
In this fashion, one can imagine a toy model in which the halo of a particular
galaxy is squashed to the best fit ($q,\phi$) within $R_{excess}$ while
outside of $R_{excess}$ we have a more nearly spherical, cosmologically 
consistent NFW$_{constr}$ halo.

Figures 7 and 8 showed that regardless of the axis ratio, the mock 
$\phi=90\degr$ rotation curves came closest to the observed rotation 
curves of UGC~4325 and UGC~5750.  To confirm that the differences between the 
rotation curves derived from the simulations and from the observed galaxy 
data are really minimized at $\phi\approx 90\degr$ and not somewhere 
between $\phi=60\degr$ and 90$\degr$, we ran additional simulations at 
$\phi=75\degr$, 85$\degr$, 86$\degr$, 87$\degr$, 88$\degr$, and 89$\degr$. 
 We then determined for each combination of $q$ and $\phi$ how well, as 
measured by $\chi^{2}_{r}$, the mock and observed rotation curves matched 
out to $R_{excess}$.  

In Figure 10, we plot the best $\phi$ for each value of $q$ for each galaxy.  
For nearly all of the 9 simulated galaxies, the mock rotation curves are the 
most consistent with the \Dpak\ galaxy rotation curves when $\phi$ is 
between 85$\degr$ and 90$\degr$ for all values of $q$.  A 
$\phi \rightarrow 90\degr$ means that the elongated axis of the NFW 
potential is pointing perpendicular to our line-of-sight.  This required 
$\phi$ is completely inconsistent with a random distribution of halo 
orientations on the sky. 

There is one galaxy in our sample, F563-V2, which has a preferred value of 
$\phi$ \textit{other} than 90$\degr$.  This galaxy has a bar in it 
\citep{pildis}.  We ran additional simulations for F563-V2 found that 
$\phi \approx 55\degr$ is the optimal viewing angle for matching the mock 
rotation curves to the observed rotation curves.  This position angle matches 
that of the bar.  It would therefore seem that we have detected the expected 
noncircular motion associated with the bar rather than the squashing of the 
halo \citep[see also][]{Spekkens07}.  

This result for F563-V2 confirms that we \textit{are} able to detect the 
presence and orientation of an asymmetry in a velocity field.  If the other 
\Dpak\ galaxies contain bars or are embedded in non-spherical NFW halos, we 
would be able to detect the asymmetry.  That $\phi \approx 90\degr$ for all 
the other \Dpak\ galaxies demonstrates that either the data are inconsistent 
with non-axisymmetric NFW halos, or that we must accept the unlikely 
coincidence that all of these galaxies are oriented such that the elongated 
axis of the potential is perpendicular to our line-of-sight.  It is not 
surprising that the effect goes in this sense as the rotation curves of LSB 
galaxies are persistently measured to be shallower than expected for NFW halos.

There are two galaxies whose results are not shown in Figure 10: F563-1 and 
NGC~4395.  The value of $\phi$ is unconstrained for both of these galaxies 
due to the radial extent of the data.  For F563-1, there are only a few data 
points to compare between the observed and mock rotation curves.  NGC~4395 
is a very nearby galaxy ($D\approx 3.5$ Mpc), and although there are many 
data points to compare in the rotation curves, the data probe a radius of less 
than $\sim$800~pc.

As discussed in \S~4.2, $q$ and $\phi$ can be used to change the amplitude of 
the NFW rotation curve.  They do not, however, alter the overall shape of 
that rotation curve.  This is reflected by high $\chi^{2}_{r}$ values for 
the comparisons of the mock and observed rotation curves.  Although the 
$\chi^{2}_{r}$ values are typically greater than 1, the sharp decline in 
$\chi^{2}_{r}$ as $\phi \rightarrow 90\degr$ indicates that 
$\phi \sim 90\degr$ is truly the minimum (see Figure 11), even though the mock 
rotation curves are not formally acceptable fits to the observed data.  
In Figure 12 we plot the ``best-fitting'' $\phi \rightarrow 90\degr$ rotation 
curves ($\phi \rightarrow 55\degr$ for F563-V2) over the observed galaxy data 
for each galaxy.  The two galaxies with the highest spatial resolution, 
DDO~64 and UGC~4325, are clear examples of how the mock rotation curve has 
shifted down in velocity such that the inner half of the mock rotation curve 
is roughly consistent with the observed data, but the outer half of the mock 
rotation curve falls below the observed data.  

Despite not being able to fully match the observed galaxy rotation curves 
within $R_{excess}$, the important trend in Figures 10 and 11, and reinforced 
by Figures 4-6, is that $\phi$ is being driven toward 90$\degr$ if one wants 
to match the observed data at small radii where the cusp/core problem is most 
severe.  This means that the elongated axis of the NFW potentials for every 
\Dpak\ galaxy (with the exception of the barred galaxy F563-V2) must point 
perpendicular to our line-of-sight.  This required $\phi$ is completely 
inconsistent with a random distribution of halo orientations on the sky.  A 
non-axisymmetric potential with a fixed axis ratio may be able to bring 
\textit{parts} of the NFW rotation curve into agreement with the observed 
data for individual galaxies, but in general, very peculiar, 
observer-dependent conditions must occur.

It is worth mentioning here that the mismatches between the ``best-fitting'' 
mock rotation curves and the data, as seen in Figure 12 for example, are a 
result of real differences in the velocity fields; information is not being 
lost or suppressed as ROTCUR collapses all the data contained in the 
two-dimensional velocity fields into a one-dimensional representation of the 
rotation.  As was shown in Figures 4 and 5, the lowest velocity portions of a 
particle's orbit are being preferentially observed in  the  $\phi=90\degr$ 
viewing orientation, ensuring that the observed range of velocities detected 
in the mock velocity fields is both small and slowly varying.  But like the 
rotation curves show, only some parts and not all of the observed and mock 
data can be made to match.  Shown as an example in Figure 13 are the observed 
UGC~4325 \Dpak\ velocity field, the ``best-fitting'' $\phi=90\degr$ 
velocity field, and the residual velocity field showing the differences 
between the two.  While the residuals are relatively small in the central 
regions of the velocity field, there are multiple areas at larger radii where 
the residuals are in the range of $\sim$ 10-15 \kms, or more.  As was 
discussed in \S~3, in galaxies where the fiber-to-fiber velocity dispersions 
are measured to be $\sim$ 6-10 \kms, 15\kms\ residuals imply a significant
mismatch.

\section{Discussion and Conclusions}
In this paper we have simulated the two-dimensional NFW halo and
tested several modifications to the potential in an attempt to
simultaneously reconcile both the NFW velocity field and rotation
curve with observed \Dpak\ galaxy data.  We have found that it is
difficult to make the cuspy NFW halo appear consistent with core-like
data without violating the predicted range of NFW parameters expected
in the $\Lambda$CDM cosmological model. 

Beginning with simulations of an axisymmetric NFW potential, we found that 
both the \Dpak\ IFU instrument and the rotation curve fitting program ROTCUR 
are able to successfully identify the NFW potential.  Observed galaxy data is 
inconsistent with the velocity fields and rotation curves corresponding to 
axisymmetric NFW potentials.  The simulated observations show that our data 
would detect the NFW cusp if it were present.  

We also tried a non-axisymmetric potential with a fixed axis ratio.  We found 
that if the parameters of the NFW halo (determined from the $c-V_{200}$ 
relation) are held constant and only the axis ratio and viewing orientation 
are varied, parts of the mock velocity fields and mock rotation curves, but 
not their entire area or length, could be made to roughly match the observed 
galaxy data.  The axis ratio and viewing orientation work to 
\textit{change only the normalization, not the radial behavior,} 
of the mock data.  The shape of the predicted NFW rotation curve remains 
distinct from the observations.  This remains true even if a more slowly 
rising NFW rotation curve is simulated. 

Only when the elongated axis of the mock galaxies is oriented perpendicular 
to the observer's line-of-sight ($\phi \sim 90\degr$) are the critical 
velocities at small radii in both the velocity fields and rotation curves 
consistent with observations.  Having all halos elongated perpendicular to 
our line of sight is clearly not reasonable.  This constraint on $\phi$ can 
be relaxed if we start with an NFW halo having a lower $V_{200}$ (see Figure 
9), \textit{but changing the halo parameters requires that we disregard 
cosmological constraints.}  Numerical simulations have shown that scatter in 
the $c-V_{200}$ relation exists ($\Delta$($\log$ c) = 0.18; 
\citet{Bullock01}), but this is not enough to ``fix'' the mock NFW data, as 
the values of the halo parameters would have to be outside the range of the 
allowable dispersion.  Furthermore, even though lowering $V_{200}$ allows for 
more scatter in $\phi$, the problem with the shape of the mock NFW rotation 
curve remains unresolved. 

In order to reconcile both the entire area of the NFW mock velocity fields and 
the entire length of their derived rotation curves with galaxy data, we need 
an asymmetry that preferentially suppresses velocities at small radii.  The 
asymmetric NFW potentials with fixed axis ratios that we have tested in this 
paper have altered the velocities at \textit{all} radii, either suppressing 
all the measured velocities (the $\phi=90\degr$ case) or boosting all of 
them (the $\phi \rightarrow 0\degr$ cases).     One possible way to address that 
problem may be to invoke a non-axisymmetry that varies with radius 
\citep{Hayashi07}.  \citet{Hayashi07} have suggested that galaxy-sized CDM 
halos are triaxial with radially varying axis ratios.  They find the halo 
potential to be highly elongated near the center ($b$/$a$ $\rightarrow$ 0.78 
and $c$/$a$ $\rightarrow$ 0.72) and increasingly more spherical at large 
radii.  This is the general behavior that the results of our two-dimensional 
simulations suggest is required.  

It remains to be seen though if such an asymmetry is a viable solution to the 
problem.  The simulations will always be constrained by the fact that LSB 
galaxy velocity fields and rotation curves are slowly rising.  This will lead 
to the problem of a preferential viewing angle that we have already 
encountered. Any potential that deviates significantly from axisymmetry at the 
center will have to be viewed at an angle which lowers, not increases, the 
observed velocities; the inner ellipsoid will need to be perpendicular to the 
line-of-sight.  On the other hand, if one insists that a radially-varying 
asymmetry \textit{is} the correct solution and that the viewing orientation 
is randomly distributed, both slowly and rapidly rising LSB galaxy rotation 
curves should be observed.  LSB galaxy rotation curves that are steeper than 
NFW rotation curves are not generally found.  Of the 50+ long-slit rotation 
curves in the literature \citep[e.g.][]{Zackrisson06,Spekkens05,dBB,MRdB}, 
it is very common to see slowly rising rotation curves, but exceedingly rare 
to see rotation curves that are in excess of the expectation for NFW halos 
that obey the $\Lambda$CDM $c-V_{200}$ relation.

The analysis of an NFW potential with a variable axis ratio is sufficiently 
complex that if one is to do it right, it should be done in three dimensions 
with interacting particles.  If having a highly elongated potential is key to 
``fixing'' the NFW velocity field and rotation curve, then adiabatic 
contraction should also be considered, as it will round out the inner halo 
potential \citep{Dubinski94,Bailin07}.  Adiabatic contraction also has the 
effect of increasing the dark matter density of the halo over its initial 
value and serves to worsen the concentration problem 
\citep[e.g.,][]{Gnedin04, SellwoodMcGaugh}.  We therefore defer the analysis 
of a radially varying axis ratio to a forthcoming paper in which the 
complexity of the simulations is increased by moving to three dimensions and 
including gas physics.

\acknowledgements
We thank the referee for a thorough, detailed, and constructive report.    
The work of R.~K.~D. and S.~S.~M. was supported by NSF grant 
AST0505956. R.~K.~D. is also supported by an NSF Astronomy \&
Astrophysics Postdoctoral Fellowship under award AST0702496.  
This paper was part of R.~K.~D.~'s Ph.~D. dissertation at 
the University of Maryland. J.~C.~M. is supported by NSF grants AST0607526 
and AST0707793.


\begin{thebibliography}{}
\bibitem[Bailin et al.(2007)]{Bailin07} Bailin, J., Simon, J.~D., Bolatto, A.~D., Gibson, B.~K., \& Power, C. 2007, \apj, 667, 191
\bibitem[Begeman(1989)]{Begeman} Begeman, K. 1989, \aap, 223, 47
\bibitem[Bolatto et al.(2002)]{Bolatto02} Bolatto, A.~D., Simon, J.~D., Leroy, A., \& Blitz, L. 2002, \apj, 565, 238
\bibitem[Borriello \& Salucci(2001)]{Borriello} Borriello, A., \& Salucci,
P. 2001, \mnras, 323, 285
\bibitem[Bosma(1981)]{Bosma} Bosma, A. 1981, \aj, 86, 1825
\bibitem[Bullock et al.(2001)]{Bullock01} Bullock, J.~S., Kolatt, T.~S., Sigad, Y., Somerville, R.~S., Kravtsov, A.~V., Klypin, A.~A., Primack, J.~R., \& Dekel, A. 2001, \mnras, 321, 559
\bibitem[Chemin et al.(2004)]{Chemin04} Chemin, L., et al. 2004, IAUS, 220, 333
\bibitem[C\^{o}t\'{e}, Carignan, \& Freeman(2000)]{Cote} C\^{o}t\'{e}, S.,
Carignan, C., \& Freeman, K.C. 2000, \aj, 120, 3027
\bibitem[de Blok \& Bosma(2002)]{dBB} de Blok, W.~J.~G., \& Bosma, A. 
2002, \aap, 385, 816 
\bibitem[de Blok et al.(2003)de Blok, Bosma, \& McGaugh]{dBBM} de Blok, W.~J.~G., Bosma, A., \& McGaugh, S.~S. 2003, \mnras, 340, 657
\bibitem[de Blok \& McGaugh(1996)]{dBM96} de Blok, W.~J.~G., \& McGaugh,
S.~S. 1996, \apj, 469, L89
\bibitem[de Blok \& McGaugh(1997)]{dBM97} ---------- . 1997, \mnras, 290, 533
\bibitem[de Blok et al.(2001)de Blok, McGaugh, \& Rubin] {dBMR} de Blok, W.~J.~G., 
McGaugh, S.~S., \& Rubin, V.~C. 2001, \aj, 122, 2396 
\bibitem[de Blok et al.(1996)de Blok, McGaugh, \& van der Hulst]{DMV} de Blok,
W.~J.~G., McGaugh, S.~S., \& van der Hulst, J.~M. 1996, \mnras, 283, 18

\bibitem[Diemand et al.(2005)]{Diemand} Diemand, J., Zemp, M., Moore, B.,
Stadel, J., \& Carollo, M. 2005, \mnras, 364, 665
\bibitem[Dubinski(1994)]{Dubinski94} Dubinski, J. 1994, ApJ, 431, 617
\bibitem[Flores \& Primack(1994)]{Flores} Flores, R.~A., \& Primack, J.~R.
1994, \apjl, 427, L1

\bibitem[Fuchs(2003)]{Fuchs} Fuchs, B. 2003, \apss, 284, 719
\bibitem[Gentile et al.(2005)]{Gentile05} Gentile, G., Burkert, A.,
  Salucci, P., Klein, U., \& Walter, F. 2005, \apj, 634, L145
\bibitem[Gentile et al.(2007)]{Gentile07} Gentile, G., Salucci, P.,
  Klein, U., \& Granato, G.~L. 2007, \mnras, 375, 199
\bibitem[Gnedin et al.(2004)]{Gnedin04} Gnedin, O.~Y., Kravtsov, A.~V.,
  Klypin, A.~A., \& Nagai, D. 2004, \apj, 616, 16
\bibitem[Hayashi et al.(2007)Hayashi, Navarro, \& Springel]{Hayashi07} Hayashi, E.,
  Navarro, J.~F., \& Springel, V.  2007, \mnras, 377, 50
\bibitem[Jing(2000)]{Jing} Jing, Y. 2000, \apj, 535, 30
\bibitem[Kuzio de Naray et al.(2006)]{K06} Kuzio de Naray, R.,
  McGaugh, S.~S., de Blok, W.~J.~G., \& Bosma, A. 2006, \apjs, 165, 461 (K06)
\bibitem[Kuzio de Naray et al.(2008)Kuzio de Naray, McGaugh, \& de Blok]{K08} Kuzio de Naray, R., McGaugh, S.~S., \& de Blok, W.~J.~G. 2008, \apj, 676, 920 (K08)
\bibitem[Marchesini et al.(2002)]{Marchesini} Marchesini, D.,
D'Onghia, E., Chincarini, G., Firmani, C., Conconi, P., Molinari, E.,
\& Zacchei, A. 2002, \apj, 575, 801
\bibitem[McGaugh et al.(2003)]{McGaugh03} McGaugh, S.~S., Barker, M.~K., \& 
de Blok, W.~J.~G. 2003, \apj, 584, 566
\bibitem[McGaugh et al.(2007)]{McGaugh07} McGaugh, S.~S., de Blok, W.~J.~G., Schombert, J.~S., Kuzio de Naray, R., \& Kim, J.~H. 2007, \apj, 659, 149
\bibitem[McGaugh, Rubin, \& de Blok(2001)]{MRdB} McGaugh, S.~S., Rubin, 
V.~C., \& de Blok, W.~J.~G. 2001, \aj, 122, 2381  
\bibitem[Mihos, McGaugh, \& de Blok(1997)]{Mihos97} Mihos, J.~C., McGaugh, S.~S., \& de Blok, W.~J.~G. 1997, \apjl, 477, L79
\bibitem[Moore et al.(1999)]{Moore} Moore, B., Quinn, T., Governato, 
F., Stadel, J., Lake, G. 1999, \mnras, 310, 1147
\bibitem[Navarro et al.(1996)Navarro, Frenk, \& White]{NFW96} Navarro, J.~F., Frenk, 
C.~S., \& White, S.~D.~M. 1996, \apj, 462, 563
\bibitem[Navarro et al.(1997)Navarro, Frenk, \& White]{NFW97} ---------- . 1997, \apj, 490, 493
\bibitem[Navarro et al.(2004)]{Navarro2004} Navarro, J.F., et al. 2004, \mnras, 349, 1039
\bibitem[Pildis et al.(1997)]{pildis} Pildis, R.~A., 
Schombert, J.~M., \& Eder, J.~A.\ 1997, \apj, 481, 157
\bibitem[Reed et al.(2003)]{Reed} Reed, D., Gardner, J., Quinn, T., 
Stadel, J., Fardal, M., Lake, G., \& Governato, F.  2003, \mnras, 346, 565
\bibitem[Rubin et al.(1980)]{Rubin} Rubin, V.~C., Thonnard, N., \& Ford, W.~K., Jr. 1980, \apj, 238, 471
\bibitem[Sellwood \& McGaugh(2005)]{SellwoodMcGaugh} Sellwood, J.~A.,
  \& McGaugh, S.~S. 2005, \apj, 634, 70
\bibitem[Schoenmakers, Franx, \& de Zeeuw(1997)]{Schoenmakers} Schoenmakers, R.~H.~M., Franx, M., \& de Zeeuw, P.~T. 1997, \mnras, 292, 349
\bibitem[Simon et al.(2005)]{Simon05} Simon, J.~D., Bolatto, A.~D., 
Leroy, A., Blitz, L., \& Gates, E. 2005, \apj, 621, 757
\bibitem[Spekkens, Giovanelli, \& Haynes(2005)]{Spekkens05} Spekkens, K., Giovanelli, R., \& Haynes, M.~P. 2005, \aj, 129, 2119
\bibitem[Spekkens \& Sellwood(2007)]{Spekkens07} Spekkens, K., \&
  Sellwood, J.~A. 2007, \apj, in press (astro-ph/0703688)
\bibitem[Swaters et al.(2003a)]{Swaters03a} Swaters, R.~A., Madore, B.~F., 
van den Bosch, F.~C., \& Balcells, M. 2003a, \apj, 583, 732
\bibitem[Swaters et al.(2003b)]{Swaters03b} Swaters, R.~A., Verheijen, 
M.~A.~W., Bershady, M.~A., \& Andersen, D.~R. 2003b, \apj, 587, L19
\bibitem[Tegmark et al.(2004)]{Tegmark} Tegmark, M., et al. 2004, \prd, 69, 103501
\bibitem[Teuben(1995)]{Teuben} Teuben, P.J. The Stellar Dynamics
Toolbox NEMO, in: Astronomical Data Analysis Software and Systems
IV, ed. R. Shaw, H.E. Payne and J.J.E. Hayes. (1995), PASP Conf
Series 77, p398
\bibitem[Wechsler et al.(2002)]{Wechsler} Wechsler, R.~H., Bullock, J.~S., Primack, J.~R., Kravtsov, A.~V., \& Dekel, A. 2002, \apj, 568, 52
\bibitem[Wong, Blitz, \& Bosma(2004)]{Wong} Wong, T., Blitz, L., \& Bosma, A. 2004, \apj, 605, 183
\bibitem[Zackrisson et al.(2006)]{Zackrisson06} Zackrisson, E., Bergvall, N., Marquart, T., \& \"{O}stlin, G. 2006, \aap, 452, 857
\bibitem[Zentner \& Bullock(2002)]{Zentner02} Zentner, A.~R., \& Bullock, 
J.~S. 2002, \prd, 66, 043003
\end{thebibliography}
\end{document}